\RequirePackage{silence}
\documentclass[aps,prb,reprint,superscriptaddress,floatfix,longbibliography]{revtex4-2}
\usepackage[T1]{fontenc}
\usepackage[utf8]{inputenc}
\usepackage{lmodern,amsmath,amssymb,bm,graphicx,microtype}
\usepackage[colorlinks=true,linkcolor=blue,citecolor=blue,urlcolor=blue]{hyperref}

\graphicspath{{./}{figures/}}
\DeclareMathOperator{\sgn}{sgn}
\newcommand{\id}{\mathbb{I}}

\allowdisplaybreaks[1]
\begin{document}

\title{Majorana interface states in anisotropic and tilted Dirac and Weyl systems}
\author{Abraham K. P. P. P\'aez}
\affiliation{Centro Universitario de los Valles, Carretera Guadalajara-Ameca Km 45.5, C.P. 46600, Ameca, Jalisco, M\'exico}
\affiliation{Tecnologico de Monterrey, School of Engineering and Sciences, Av. General Ram\'on Corona 2514, Zapopan 45138, M\'exico}
\author{Alireza Qaiumzadeh}
\affiliation{Center for Quantum Spintronics, Department of Physics, Norwegian University of Science and Technology, NO-7491 Trondheim, Norway}

\begin{abstract}
Anisotropy and cone tilt change how Majorana states propagate along an interface and decay away from it. We derive a two-dimensional Majorana surface Hamiltonian from a fully gapped three-dimensional superconductor, with the surface mass generated by the relative phase between triplet and singlet pairing. For a mass domain wall on this surface, we obtain the wave function, propagation velocity, and localization length of the resulting one-dimensional channel for a general invertible velocity tensor and an arbitrary wall orientation. The product of propagation speed and localization length is independent of wall orientation in the anisotropic model. For a smooth closed channel without a vortex, the propagation time around the wall determines the lowest excitation energy. For a tilted surface cone, the normal and parallel components of tilt affect confinement and propagation differently. We also solve a separate three-dimensional Weyl model with uniform singlet pairing and a reversing exchange field. In the regime supporting interface states, its bulk remains gapless. At zero chemical potential, two states confined perpendicular to the wall and propagating within its plane have the same dispersion over a finite momentum interval. Numerical diagonalization confirms the analytical energies and wave functions.
\end{abstract}

\maketitle

\section{Introduction}
Fully gapped topological superconductors support boundary excitations whose existence follows from the topology of the bulk quasiparticle states \cite{ReadGreen,QiZhang,xiaozhang,Kane2010,Schnyder2008}. Majorana states can occur at vortices, defects, and interfaces where the superconducting or magnetic order changes \cite{JackiwRebbi,FuKane}. Spatially separated Majorana zero modes have attracted interest for their non-Abelian exchange statistics and possible use in topological quantum computation \cite{Nayak,QiZhang,AYuKitaev_2001,KITAEV20032}. Here we study two interface geometries. A mass wall on
the two-dimensional (2D) surface of a fully gapped
three-dimensional (3D) superconductor supports a
one-dimensional Majorana channel. We also study
2D states at a planar exchange wall in a
separate 3D Weyl superconducting model, whose bulk remains
gapless in the regime supporting these interface states.

Dirac and Weyl materials connect superconductivity with spin-momentum locking and electronic band topology \cite{Armitage}. Band inversion in quantum wells \cite{Andrei2006} and Weyl nodes in 3D solids \cite{Wan} provide two examples. The superconducting spectrum depends on the pairing symmetry and on which electronic states form Cooper pairs \cite{FuKane,MengBalents,BednikZyuzinBurkov}. Experimental studies have reported unconventional superconductivity at point contacts on Cd$_3$As$_2$ \cite{Aggarwal2016} and evidence for vortex-bound Majorana states in an iron-based superconductor \cite{Dongfei2018}. Anisotropic velocities and tilted cones also occur in Dirac and Weyl materials \cite{Goerbig,Soluyanov,Deng2016,Noh2017,Li2017}. These features change the velocities entering the interface equation and thereby affect both propagation along a wall and confinement perpendicular to it.

Previous work has examined Majorana states in anisotropic systems and the dependence of their existence on interface orientation \cite{mendler2017,Vasenko2017}. We formulate this dependence in terms of a general velocity tensor. For any straight wall and an invertible tensor, we derive the normalized wave function, propagation velocity, and localization length. The determinant of the tensor fixes the propagation direction for a specified mass wall, while its components and the wall angle determine the speed and confinement. We also obtain the finite-size excitation energy for a channel with antiperiodic boundary conditions and express the corresponding result for a curved closed wall through the propagation time around it.

We then include cone tilt in the 2D Majorana surface Hamiltonian. The tilt leaves the bulk spinors at fixed momentum unchanged, but modifies the interface spinor, propagation velocity, and localization length. Resolving the tilt into components parallel and perpendicular to the wall gives the solution at an arbitrary angle. When the normal is parallel to the tilt, the propagation speed, localization length, and lowest finite-size excitation energy all decrease as the tilt approaches its critical value.

In 3D, we first evaluate the projection of spin-singlet pairing onto a Weyl band \cite{FuKane,MengBalents,BednikZyuzinBurkov,Haldane2018}. We then solve a spinful Bogoliubov--de Gennes (BdG) model with uniform singlet pairing and an exchange field that reverses across a plane. At zero chemical potential, we obtain two interface states with equal dispersion, their decay lengths on both sides, and the momentum interval in which they are localized. These results connect with studies of tilt-dependent superconducting nodes and Majorana bands \cite{Alidoust}, and of proximity-induced Weyl-surface states and their Josephson response \cite{chen2016weyl,chen2017josephson}.

The surface mass has a direct superconducting interpretation. Qi, Witten, and Zhang showed that a relative superconducting phase can give a mass to a Majorana surface cone of a gapped 3D parent \cite{xiaozhang}. Their axion phase theory describes dynamical condensate phases, while the corresponding parameter in a topological insulator is fixed in the bulk \cite{QiHughesZhang}. Electromagnetic response in a superconductor also involves the condensate and its coupling to the gauge field \cite{Masatoshi2021}. Topological contributions to the superfluid stiffness and
Josephson quantum capacitance have also been studied in
superconducting Dirac systems \cite{Wang2024}, while the axionic
response associated with charge-density-wave order has been
studied in Weyl semimetals \cite{Roy2015}. We state the phase and Chern-number conventions explicitly and derive the Majorana surface mass from a spinful superconducting parent.

Section~\ref{Anis} treats the anisotropic surface, its mass wall, and finite-size excitation energies. Section~\ref{Tilt} derives the tilted-wall solution. Section~\ref{Weyl} treats Weyl-band pairing and the exchange wall, and Section~\ref{summ} summarizes the results. Appendix~\ref{app:parent} derives the surface Hamiltonian from a gapped parent. Appendix~\ref{app:numerics} compares the analytical solutions with numerical diagonalization and reports convergence with grid spacing and box size.

\section{Majorana States at Anisotropic Surface Mass Walls} \label{Anis}
We begin by describing quasiparticles on the 2D surface of a gapped 3D topological superconductor. In the Majorana representation, the field satisfies $\gamma_a(\bm k)^\dagger=\gamma_a(-\bm k)$, and the quadratic Hamiltonian is $\mathcal H=\frac12\int d^2r\,\gamma^{\mathrm T}H\gamma$. The Pauli matrices below act on the two Majorana surface components. The corresponding electron and hole states are specified by the parent model. An explicit four-component parent construction and surface projection are given in Appendix~\ref{app:parent}. To account for general linear anisotropy, we introduce a real velocity tensor with $\det\bm V\ne0$ and use units $\hbar=1$:

\begin{equation}\label{H_0}
    H_0(\bm{k})=\bm{d}(\bm{k})\cdot\bm{\sigma},
\end{equation}
where $\bm{d}(\bm{k})=(d_{x},0,d_z)=(v_{xx}k_{x}+v_{yx}k_y, 0, v_{xz}k_{x}+v_{yz}k_y)$ and $\bm{\sigma}=(\sigma_{x},\sigma_y,\sigma_z)$ denote Pauli matrices in the Majorana surface subspace. In the absence of a mass term, the Hamiltonian is invariant under time-reversal symmetry, $\mathcal{T} H_0(\bm{k})\mathcal{T} ^{-1}=H_0(-\bm{k})$, with $\mathcal T=i\sigma_y\mathcal K$ and $\mathcal T^2=-1$, where $\mathcal K$ denotes complex conjugation.  The corresponding energy spectrum is
\begin{equation}\label{E_0}
    E_{\pm}(\bm{k})=\pm\sqrt{(v_{xx}k_{x}+v_{yx}k_y)^2+(v_{xz}k_{x}+v_{yz}k_y)^2},
\end{equation}
which describes a gapless anisotropic 2D Dirac cone, where the positive and negative branches are related by the intrinsic particle-hole redundancy of the Majorana field. In contrast to the isotropic case, the quasiparticle velocity depends explicitly on the propagation direction through the Fermi velocity tensor $v_{ij}$.

\begin{figure*}[t]
\centering\includegraphics[width=\linewidth]{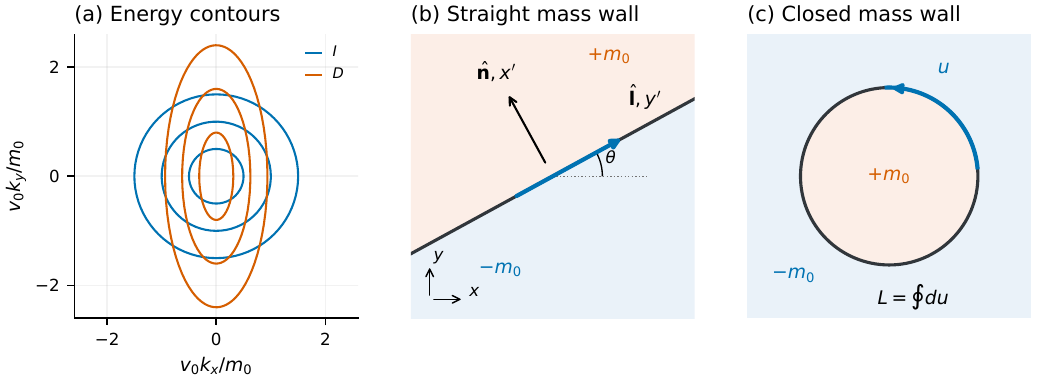}
\caption{Energy contours and mass-wall geometry. (a) Positive-energy contours $E/m_0=0.5,1,1.5$ for the massless Hamiltonian with isotropic velocity tensor $\bm V=v_0I$ and anisotropic velocity tensor $\bm V=v_0D$, where $I$ is the identity matrix and $D=\operatorname{diag}(1.6,0.625)$. Momentum is measured in units $m_0/v_0$. The two tensors have the same determinant and different energy contours. (b) Straight wall: the mass changes from $-m_0$ to $+m_0$ along the normal $\hat{\bm n}$. The tangent $\hat{\bm l}$ makes an angle $\theta$ with the $x$ axis. The blue arrow shows propagation for $\det\bm V>0$. (c) Closed wall with positive mass inside and negative mass outside. The blue arrow gives the direction of increasing arclength $u$. Equation~\eqref{loopgap} relates the lowest excitation energy, for antiperiodic boundary conditions, to the propagation time around this wall.}\label{geometry}
\end{figure*}

Figure~\ref{geometry}(a) compares the isotropic velocity tensor
$\bm V=v_0I$ with the anisotropic example $\bm V=v_0D$, where
$v_0>0$ is a reference velocity, $I$ is the identity matrix,
and $D=\operatorname{diag}(1.6,0.625)$ is dimensionless.
The two tensors have the same determinant but different
matrices $\bm V^{\mathrm T}\bm V$, and therefore different
constant-energy contours. For a fixed tensor, changing the
energy rescales the contours without changing their shape.

To generate localized boundary states, we introduce a mass term \cite{JackiwRebbi, Kane2010, Haldane} to the Hamiltonian,
\begin{equation}
    H_m=m(\bm{r})\sigma_y,
\end{equation}
which breaks time-reversal symmetry and opens a gap in the 2D Dirac spectrum. The full Hamiltonian $H_{\rm BdG}=H_0+H_m$ describes a surface Majorana cone. In the construction of Qi, Witten, and Zhang, two conventional superconducting films with phases $+\pi/2$ and $-\pi/2$ relative to the gapped parent induce opposite surface masses \cite{xiaozhang}. The parameter $m$ is the surface mass induced by this relative superconducting phase.

The particle-hole operator is $\mathcal C=\mathcal K$, and $\mathcal C H_{\rm BdG}(\bm k)\mathcal C^{-1}=-H_{\rm BdG}(-\bm k)$ with $\mathcal C^2=+1$. A nonzero mass breaks $\mathcal T$ and places the surface in class D. The field constraint and this matrix identity relate the positive- and negative-energy solutions through $E_-(\bm k)=-E_+(-\bm k)$ \cite{QiZhang,Schnyder2008}. They are the particle-hole partners of the same Majorana field.

For a spatially uniform mass, the spectrum is fully gapped, whereas a sign reversal of the mass across an interface gives rise to topologically protected bound states through the Jackiw--Rebbi mechanism \cite{JackiwRebbi}. 
Figures~\ref{geometry}(b) and \ref{geometry}(c) illustrate
the straight and closed interfaces formed by this mass
sign reversal.
The topology of the massive 2D Dirac Hamiltonian is characterized by the Berry curvature of the negative-energy band, using $\bm{\mathcal A}=i\langle u_-|\nabla_{\bm k}u_-\rangle$ \cite{XiaoChangNiu}. Within the continuum model, the corresponding contribution to the Chern number is given by \cite{Thouless82,Kohmoto85, QiHughesZhang} 
\begin{equation}
    C=\frac{1}{4\pi}\int d^2 \bm{k} \frac{\mathbf d\cdot(\partial_{k_{x}}\mathbf d\times\partial_{k_y}\mathbf d)}{|\mathbf d|^3},
\end{equation}
where the $\bm{d}$ vector now becomes
$\bm {d}=(d_{x},m,d_z)$.
Introducing the velocity matrix
\begin{equation}
    \bm{V}=\begin{pmatrix}
        v_{xx} & v_{yx}\\
        v_{xz} & v_{yz}
        \end{pmatrix},
\end{equation}
the Chern-number integral, in the presence of a uniform mass term, becomes 
\begin{equation}
    C=-\frac{1}{4\pi}\int d^2 \bm{k} \frac{m \det(\bm{V})}{|\mathbf{d}|^{3}},
\end{equation} 
where $|\mathbf{d}|=\sqrt{(v_{xx}k_{x}+v_{yx}k_y)^2+(v_{xz}k_{x}+v_{yz}k_y)^2+m^2}$. The integral is evaluated by introducing the transformed momentum
$\bm{k}'=\bm{V}\bm{k}$,
whose Jacobian is
$d^2 \bm{k}={d^2\bm{k}'}/{|\det(\bm{V})|}$.
This yields
\begin{equation}
    C=-\frac{\sgn[\det(\bm{V})]}{4\pi}\int d^2k' \frac{m}{(k'^2+m^2)^{3/2}},
\end{equation}
and therefore
\begin{equation}\label{chern}
    C=-\frac{1}{2} \sgn(m) \sgn[\det(\bm{V})].
\end{equation}
The spectrum in Eq.~\eqref{E_0} depends on $|\bm V\bm k|^2$. In particular, $\bm V$ and $O\bm V$ have identical spectra for an orthogonal matrix $O$, while a reflection with $\det O=-1$ reverses the sign in Eq.~\eqref{chern}. Thus equal matrix products $\bm V^{\mathrm T}\bm V$ give equal dispersions, while the sign of the determinant can distinguish their propagation directions at a fixed mass wall.

The half-integer value in Eq.~\eqref{chern} is the contribution of one massive continuum cone, commonly called the parity anomaly \cite{Haldane,QiZhang,QiHughesZhang,Redlich1984,Niemi1983}. For two mass domains with the same large-momentum regularization, their difference is an integer. With the normal, tangent, and mass profile defined below, this difference is $\Delta C=C(x'<0)-C(x'>0)=\sgn\det\bm V$. Its magnitude gives one interface channel, and its sign fixes the propagation direction \cite{JackiwRebbi,xiaozhang}. The cancellation between the boundary contribution of the bulk theory and that of the interface state is the anomaly-inflow relation \cite{Callan1985}.

The axion phase theory connects the surface mass to a gapped 3D superconductor. In the conventions of Qi, Witten, and Zhang \cite{xiaozhang}, its action contains
\begin{equation}\label{topological}
 \mathcal L_{\rm topo}=\sum_i\frac{C_{1i}\vartheta_i}{64\pi^2}
 \epsilon^{\mu\nu\sigma\tau}F_{\mu\nu}F_{\sigma\tau}.
\end{equation}
Here $C_{1i}$ is the integer Chern number of a closed normal-state \emph{3D Fermi surface}, and $\vartheta_i$ is its pairing phase. We retain the source convention $a_i=-i\langle u_i|\nabla_{\bm k}u_i\rangle$ and the orientation assigned there to each Fermi surface. The integer $C_{1i}$ already contains the orientation information entering this bulk coefficient. In charge units $e=1$, the phase action in the convention
of Ref.~\cite{xiaozhang} can be written as
\begin{equation}\label{phaseaction}
\begin{aligned}
 \mathcal L_{\rm eff}={}&\mathcal L_{\rm Maxwell}+\mathcal L_{\rm topo}\\
 &+\frac12\sum_i\rho_i^{\mu\nu}
 (\partial_\mu\vartheta_i-2A_\mu)(\partial_\nu\vartheta_i-2A_\nu)\\
 &+\sum_{i<j}J_{ij}\cos(\vartheta_i-\vartheta_j).
\end{aligned}
\end{equation}
Here $A_\mu$ is the electromagnetic four-potential,
$F_{\mu\nu}=\partial_\mu A_\nu-\partial_\nu A_\mu$ is its
field tensor, and $\mathcal L_{\rm Maxwell}$ is the Maxwell
Lagrangian density. The coefficients $\rho_i^{\mu\nu}$
describe the compressibility and superfluid stiffness of
condensate $i$, and $J_{ij}$ is the Josephson coupling
between condensates $i$ and $j$. Repeated Greek indices
are summed over time and the three spatial coordinates;
$\epsilon^{\mu\nu\sigma\tau}$ is the totally antisymmetric
symbol.

Under an electromagnetic gauge transformation, $A_\mu\mapsto A_\mu+\partial_\mu\chi$ and $\vartheta_i\mapsto\vartheta_i+2\chi$, the phase gradients in Eq.~\eqref{phaseaction} are invariant. For the complete set of bulk Fermi surfaces, $\sum_i C_{1i}=0$, so the combination $\Theta=\sum_i C_{1i}\vartheta_i$ in Eq.~\eqref{topological} is also invariant: it depends on relative condensate phases. Gauge invariance therefore uses the complete set of bulk Fermi surfaces. For a fully gapped, time-reversal-invariant superconductor in the weak-pairing regime, the bulk invariant is
\begin{equation}\label{parentinvariant}
 N=\frac12\sum_i C_{1i}\sgn\delta_i,
\end{equation}
where $\delta_i$ is the real pairing amplitude in a fixed time-reversal convention, and the Fermi surfaces are taken to be nondegenerate \cite{QHZ2010,xiaozhang}. For a spin-degenerate normal state such as the untilted parent in Appendix~\ref{app:parent}, one can introduce an arbitrarily small time-reversal-invariant spin-orbit splitting without closing the superconducting gap before applying this formula \cite{xiaozhang}. The index $N$ describes the 3D bulk, while Eq.~\eqref{chern} describes a massive surface cone. Equation~\eqref{topological} uses the original normalization of Ref.~\cite{xiaozhang}. St\aa{}lhammar \emph{et al.}~\cite{Masatoshi2021} obtain an electromagnetic topological coefficient with one third of that magnitude in the two-helicity model, after matching conventions. Their calculation includes condensate dynamics and the boundary states, and gives zero electric Hall response for a time-reversal-invariant interface between a topological and a trivial superconductor. The electric response and the surface-cone Chern number therefore describe different observables. For the wall considered here, the required physical connection is the mass generated by the relative superconducting phase. Appendix~\ref{app:parent} derives this mass by projecting the parent Hamiltonian onto its surface states.

To determine the corresponding bound-state wave function, we consider a straight domain wall oriented along the unit vector
\begin{equation}\label{l-def}
    \hat{\bm{l}}=(\cos\theta,\sin\theta),
\end{equation}
with normal direction
\begin{equation}\label{n-def}
    \hat{\bm{n}}=(-\sin\theta,\cos\theta).
\end{equation}
The mass profile is assumed to change sign across the interface \cite{xiaozhang},
\begin{equation} \label{mass}
    m(x')=m_0 \sgn(x'),
\end{equation}
where $m_0>0$ and $x'$ denote the coordinates measured along $\hat{\bm{n}}$. The bound state is therefore localized at $x'=0$, while propagating freely along the interface direction $\hat{\bm{l}}$.

The momentum components in the rotated coordinate system are related to the laboratory frame through
\begin{equation}\label{lab-frame}
    \begin{aligned}
        k_{x} &= -\sin\theta k_{x'}+\cos\theta k_{y'},\\
        k_y &= \cos\theta k_{x'}+\sin\theta k_{y'}.
    \end{aligned}
\end{equation}
Substituting these expressions into Eq. (\ref{H_0}), and introducing the vectors $\mathbf a=(v_{xx},v_{yx})$ and $\mathbf b=(v_{xz},v_{yz}),$ along with the coefficients
\begin{equation}
    \begin{aligned}
        \alpha_{x'}&=\mathbf a\cdot\hat{\bm{n}},\quad\alpha_{y'}=\mathbf a\cdot\hat{\bm{l}},\\
        \beta_{x'}&=\mathbf b\cdot\hat{\bm{n}},\quad\beta_{y'}=\mathbf b\cdot\hat{\bm{l}},
    \end{aligned}
\end{equation}
the kinetic surface Hamiltonian assumes the compact form
\begin{equation}
    H_0=(\alpha_{x'}k_{x'}+\alpha_{y'}k_{y'})\sigma_{x}+(\beta_{x'}k_{x'}+\beta_{y'}k_{y'})\sigma_z.
\end{equation}
For states propagating along the interface, we use the ansatz
\begin{equation}
    \psi(x',y')=e^{ik_{y'}y'}\phi(x'),
\end{equation}
which leads to the following effective 1D BdG Hamiltonian
\begin{equation}\label{H1D}
    \begin{aligned}
        H_{1D}=&-i\alpha_{x'}\sigma_{x}\partial_{x'}-i\beta_{x'}\sigma_z\partial_{x'}
        \\
        &+k_{y'}(\alpha_{y'}\sigma_{x}+\beta_{y'}\sigma_z)+m(x')\sigma_y .
    \end{aligned}
\end{equation}
We first consider the zero-energy solution at $k_{y'}=0$, for which
\begin{equation}
    H_{1D}=-i(\alpha_{x'}\sigma_{x}+\beta_{x'}\sigma_z)\partial_{x'}+m(x')\sigma_y .
\end{equation}
Defining $A\equiv\alpha_{x'}\sigma_{x}+\beta_{x'}\sigma_z$,
we obtain $\{A,\sigma_y\}=0.$ For $\lambda>0$, there is a rotation about the $y$ axis, $\mathcal{R}$, that transforms it into
\begin{equation}\label{alg-cond}
\mathcal{R} A \mathcal{R}^\dagger=\lambda\sigma_x,\qquad \mathcal{R}\sigma_y\mathcal{R}^\dagger=\sigma_y,
\end{equation}
with $\lambda=\sqrt{\alpha_{x'}^2+\beta_{x'}^2}$. Consequently, the Hamiltonian is unitarily equivalent to the standard Jackiw--Rebbi problem,
\begin{equation}
    \widetilde H_{1D}=\mathcal R H_{1D}\mathcal R^\dagger=-i\lambda\sigma_{x}\partial_{x'}+m(x')\sigma_y,
\end{equation}
whose zero-energy bound state is the Jackiw--Rebbi mode
\cite{JackiwRebbi,Jackiw-Rebbi}.
For the domain-wall profile \eqref{mass}, the rotated zero mode $\widetilde\phi_0=\mathcal R\phi_0$ satisfies
\begin{equation}\label{0mode1D}
    -i\lambda\sigma_{x}\partial_{x'}\widetilde\phi_0(x')+m(x')\sigma_y\widetilde\phi_0(x')=0.
\end{equation}
In the original spinor basis, the decaying solution obeys $(\alpha_{x'}\sigma_{x}+\beta_{x'}\sigma_{z}-i\lambda\sigma_{y})\chi=0$. An explicit solution to the $2\times2$ system yields
\begin{equation}
    \chi=\dfrac{1}{\sqrt{2\lambda(\lambda+\alpha_{x'})}}\begin{pmatrix}
        \beta_{x'}\\\lambda+\alpha_{x'}
    \end{pmatrix}
\end{equation}
for $\lambda+\alpha_{x'}>0$. When this expression is singular, one can use $\chi=(\lambda-\alpha_{x'},\beta_{x'})^{\mathrm T}/\sqrt{2\lambda(\lambda-\alpha_{x'})}$. The two expressions differ only by an overall sign where both are defined. Thus a single normalized spinor solves both sides of the wall. In the original basis, the normalized transverse wave function is
\begin{equation}\label{aniwavefunction}
    \phi_0(x')=\frac{1}{\sqrt{\xi}}\exp \left(-|x'|/\xi\right)\chi,
\end{equation}
where the localization length is
\begin{equation}\label{XiAnisotropic}
    \xi=\frac{\lambda}{m_0}.
\end{equation}
The parameter $\lambda=\sqrt{(\mathbf a \cdot \hat{\bm{n}})^2+(\mathbf b \cdot \hat{\bm{n}})^2}$ represents the effective quasiparticle velocity perpendicular to the interface. Consequently, the localization length depends not only on the anisotropy of the Dirac cone but also on the orientation of the domain wall.

To first order in momentum parallel to the interface, the zero mode acquires a linear dispersion,
\begin{equation}
    E(k_{y'})=v_\parallel k_{y'},
\end{equation}
where the propagation velocity is obtained by projecting the perturbation
$k_{y'}(\alpha_{y'}\sigma_{x}+\beta_{y'}\sigma_z)$, see Eq. (\ref{H1D}),
onto the zero-mode wave function. This gives
\begin{equation}\label{VperpAnisotropic}
    \begin{aligned}
        v_\parallel&=\frac{\alpha_{y'}\beta_{x'}-\alpha_{x'}\beta_{y'}}{\lambda}=\frac{\det(\bm{V})}{\lambda}\\
        &=\frac{v_{xx}v_{yz}-v_{xz}v_{yx}}{\sqrt{(\mathbf a \cdot \hat{\bm{n}})^2+(\mathbf b \cdot \hat{\bm{n}})^2}}.
    \end{aligned}
\end{equation}
The group velocity is $v_\parallel=\partial E/\partial k_{y'}$. A positive value gives propagation along $\hat{\bm l}$ and a negative value gives propagation in the opposite direction; the speed is $|v_\parallel|$. Equation~\eqref{VperpAnisotropic} shows that the determinant fixes this direction for the chosen mass profile, while the wall angle changes the magnitude through $\lambda$. 

The result is exact for the straight linear model if the finite-momentum wave function includes the transverse phase
\begin{equation}\label{anisfiniteq}
 \psi_q(x',y')=e^{iqy'}e^{-iq\kappa x'}\phi_0(x'),\qquad
 \kappa=\frac{(\bm V\hat{\bm n})\cdot(\bm V\hat{\bm l})}{\lambda^2}.
\end{equation}
Substitution into Eq.~\eqref{H1D} gives $E(q)=v_\parallel q$. In particular, $|v_\parallel|\xi=|\det\bm V|/m_0$, independent of orientation. The real zero-momentum spinor obeys $\mathcal C\phi_0=\phi_0$, the Majorana self-conjugacy condition. Finite-$q$ states occur as particle-hole partners at $\pm q$. Independent continuum diagonalization at nonzero $q$ checks both this energy and the spatial profile in Appendix~\ref{app:numerics}.

For an invertible velocity tensor, Eq.~\eqref{VperpAnisotropic}
gives a nonzero propagation velocity at every wall orientation.
Flat Majorana interface bands have also been obtained in
models whose kinetic blocks depend on a single independent
momentum combination \cite{MendlerThesis}.

To determine the finite-size excitation energy, identify points separated by a length $L$ along a straight channel and impose antiperiodic boundary conditions, $\gamma(y'+L)=-\gamma(y')$. Then $q_n=2\pi(n+1/2)/L$, with $n\in\mathbb Z$, and the smallest positive excitation energy is
\begin{equation}\label{MinigapAnisotropic}
 \Delta_{\rm fs}=\frac{\pi|v_\parallel|}{L}
 =\frac{\pi|\det\bm V|}{L\|\bm V\hat{\bm n}\|}.
\end{equation}
The energy $\Delta_{\rm fs}$ is the spacing from zero to the lowest positive channel level. An isolated subgap level satisfies $\Delta_{\rm fs}<E_{\rm bulk,gap}$, with its momentum inside the range of the surface theory. Antiperiodic boundary conditions describe a closed chiral Majorana channel without a vortex \cite{ReadGreen}. A vortex can change the boundary condition to periodic and allow a zero-energy level \cite{Ivanov2001}.

The isotropic limit is recovered by setting $v_{xx}=v_{yz}=v_F$ and $v_{yx}=v_{xz}=0$ in Eq.~\eqref{MinigapAnisotropic}. In this case, $\mathbf{a}=(v_{F},0)$ and $\mathbf{b}=(0,v_F)$, and it reduces to the result $\Delta_{\rm fs}={\pi v_F}/{L}$, in agreement with previous studies of isotropic topological superconductors \cite{FuKane,xiaozhang,Alicea2012,Beenakker2013}.

For a smooth closed wall whose local speed varies with arclength $u$, the Hermitian projected operator is $-i[v_\parallel(u)\partial_u+\tfrac12\partial_uv_\parallel(u)]$. Its travel time gives
\begin{equation}\label{loopgap}
 T=\oint\frac{du}{|v_\parallel(u)|},\qquad \Delta_{\rm fs}=\frac{\pi}{T}
\end{equation}
for antiperiodic boundary conditions. Here $T$ is the time to propagate once around the wall. A constant speed gives Eq.~\eqref{MinigapAnisotropic}. For a curved wall, its changing orientation enters through $v_\parallel(u)$. The local projection assumes a curvature radius much larger than the localization length.

\begin{figure*}[t]
\centering\includegraphics[width=.93\linewidth]{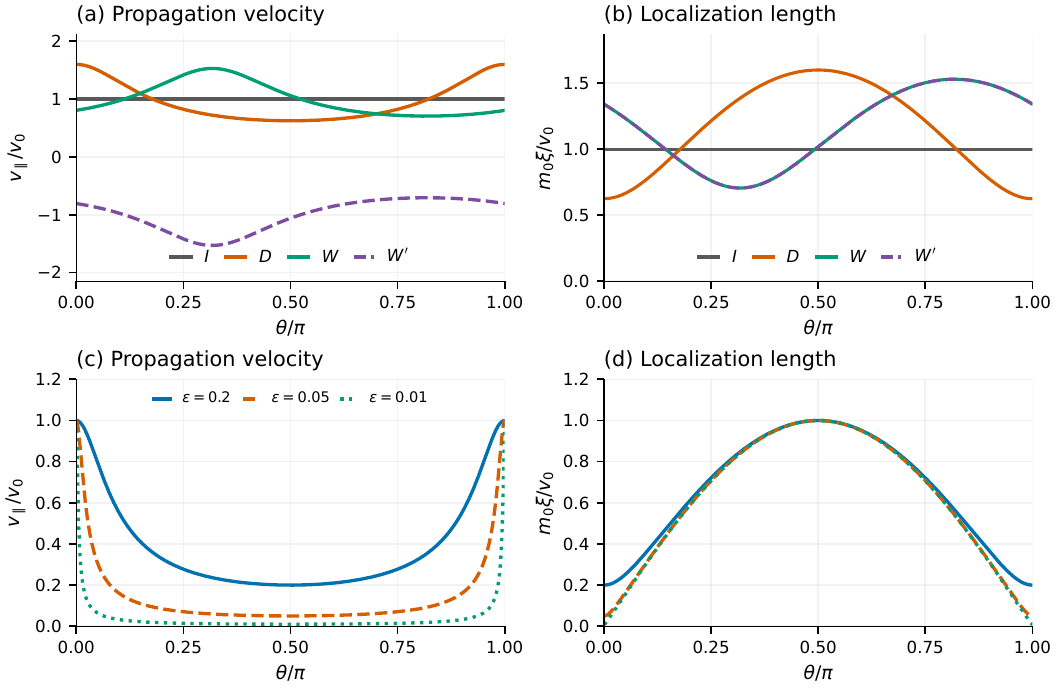}
\caption{Propagation velocity and localization length versus wall angle. Positive and negative velocities indicate motion along and opposite to $\hat{\bm l}$, respectively. (a,b) Velocity tensors in units $v_0$: $I$, $D=\operatorname{diag}(1.6,0.625)$, $W=\bigl(\begin{smallmatrix}1&0.6\\0.2&1.2\end{smallmatrix}\bigr)$, and $W'=\operatorname{diag}(-1,1)W$. The reflection from $W$ to $W'$ reverses propagation and leaves the localization length unchanged. (c,d) Tensors $\bm V/v_0=\operatorname{diag}(1,\epsilon)$ with $\epsilon=0.2,0.05,0.01$. Decreasing $\epsilon$ reduces the second principal velocity. The interface speed remains $v_0$ at $\theta=0,\pi$ and decreases toward zero at every other fixed angle. The localization length remains finite and becomes smaller at $\theta=0,\pi$. All curves have period $\pi$. For an antiperiodic channel with $L=20v_0/m_0$, $\Delta_{\rm fs}/m_0=(\pi/20)|v_\parallel/v_0|$. Curves follow Eqs.~\eqref{XiAnisotropic} and \eqref{VperpAnisotropic}.}\label{anisotropic}
\end{figure*}

Figure~\ref{anisotropic}(a,b) shows the propagation velocity and localization length as the wall angle changes. Off-diagonal tensor elements rotate the directions at which these quantities are largest and smallest. The reflection $W'=\operatorname{diag}(-1,1)W$ preserves $W^{\mathrm T}W$ and reverses the propagation direction while leaving the speed and localization length unchanged. At fixed $L$, the finite-size excitation energy follows directly from the speed through Eq.~\eqref{MinigapAnisotropic}.

Panels (c,d) show what happens when one of the two principal velocities becomes small. For $\bm V=v_0\operatorname{diag}(1,\epsilon)$, the velocities along the principal axes are $v_0$ and $\epsilon v_0$. As $\epsilon\to0^+$, the interface velocity tends to zero at every fixed angle with $\sin\theta\ne0$. At $\theta=0,\pi$, however, it remains $v_0$, while $\xi=\epsilon v_0/m_0$ tends to zero. Thus, the limiting behavior depends on wall orientation. Throughout this approach, $\xi\leq\|\bm V\|/m_0$, where $\|\bm V\|$ is the largest principal velocity. All curves are evaluated at $\epsilon>0$, as required by the invertible-tensor solution. For a microscopic realization, the inverse localization
length must also remain below the momentum cutoff of the
surface theory.

\section{Effect of Cone Tilt on Majorana Surface States}\label{Tilt}
We next include a tilt of the 2D surface cone. Tilted Dirac and Weyl dispersions occur in a variety of quantum materials and change the low-energy spectrum without changing the eigenvectors at fixed momentum \cite{Goerbig,Soluyanov}. Tilt also modifies thermoelectric transport in Weyl
semimetals \cite{PhysRevB.107.014410}. Here, we determine how tilt changes a Majorana mass-wall state.

\subsection{Hamiltonian, symmetry, and band spinors}
We consider the massless kinetic operator
\begin{equation}\label{tiltedHamiltonian}
 H_0^{(s)}(\bm k)=v_F(sk_x\sigma_x+k_y\sigma_y)
       +sv_Ft_xk_x\id,
\end{equation}
where $v_F>0$, $s=\pm1$, and the tilt vector is $\bm t^{(s)}=s t_x\hat{\bm x}$. We solve one cone at a time. The choices $s=+1$ and $s=-1$ reverse the coefficients of $k_x\sigma_x$ and the tilt term together. The surface mass is $m(x')\sigma_z$. In this basis, the particle-hole operator is $\mathcal C=\sigma_x\mathcal K$, with $\mathcal C^2=+1$ and
\begin{equation}\label{tiltPHS}
 \mathcal C H_{\rm BdG}^{(s)}(\bm k)\mathcal C^{-1}
 =-H_{\rm BdG}^{(s)}(-\bm k).
\end{equation}
The term proportional to the identity preserves particle-hole symmetry and breaks the time-reversal symmetry $\mathcal T=i\sigma_y\mathcal K$ of the massless cone. We vary the tilt while holding the mass profile at $m(x')=m_0\sgn x'$, so any uniform contribution to the mass is set to zero. Appendix~\ref{app:parent} gives a parent Hamiltonian that remains gapped under a sufficiently small tilt.

The massless spectrum is
\begin{equation}\label{energywithtilt}
 E_\pm^{(s)}(\bm k)=sv_Ft_xk_x\pm v_F\sqrt{k_x^2+k_y^2}.
\end{equation}
At charge neutrality, its zero-energy set obeys
\begin{equation}\label{DiracPoint3d}
 (t_x^2-1)k_x^2=k_y^2.
\end{equation}
For $|t_x|<1$, called type I, the only zero-energy point is $\bm k=0$. At $|t_x|=1$, the unbounded linear model has zero-energy rays along the tilt direction, and for $|t_x|>1$ it has two intersecting lines. In a material, band curvature completes the electron and hole pockets; the straight lines are their local continuum description \cite{Soluyanov}.

The normalized upper-band spinor can be written
\begin{equation}\label{diracspinor}
 |u_+^{(s)}(\bm k)\rangle=\frac{1}{\sqrt2}
 \begin{pmatrix}1\\e^{i\phi_s}\end{pmatrix},\qquad
 \phi_s=\arg(sk_x+i k_y).
\end{equation}
Thus $\phi_+=\phi$ and $\phi_-=\pi-\phi$ modulo $2\pi$, where $\phi=\arg(k_x+i k_y)$. Within either fixed sector, momentum inversion gives $\phi_s(-\bm k)=\phi_s(\bm k)+\pi$ modulo $2\pi$. With $\bm{\mathcal A}=i\langle u_+|\nabla_{\bm k}u_+\rangle$, the azimuthal vector component is $\mathcal A_\phi=-s/(2k)$ and the Berry phase is $-s\pi$ modulo $2\pi$. The tilt changes the band energy and leaves this spinor and Berry phase unchanged.

The same kinetic matrix also gives a useful electronic pairing calculation when its Pauli matrices are taken to act on electron spin. For a real spin-singlet pair potential $\hat\Delta_e=\Delta_e i\sigma_y$, the matrix element between upper-band states at opposite momenta is
\begin{equation}\label{diracpairing}
\begin{aligned}
 \Delta_{\rm eff}^{(s)}(\bm k)
 &=\langle u_+^{(s)}(\bm k)|\hat\Delta_e|u_+^{(s)*}(-\bm k)\rangle\\
 &=-\Delta_e e^{-i\phi_s}
 =\begin{cases}-\Delta_e e^{-i\phi},&s=+1,\\
                \Delta_e e^{i\phi},&s=-1.
   \end{cases}
\end{aligned}
\end{equation}
On a nondegenerate 2D Fermi contour, the projected pair amplitude acquires a phase that winds once as the momentum encircles the contour \cite{FuKane}. Its explicit phase depends on the phases chosen for the Bloch spinors. This electronic calculation uses a Nambu basis containing electron and hole states. The Majorana surface mass $m\sigma_z$, by comparison, follows from the parent-surface projection in Appendix~\ref{app:parent}. In both settings, the tilt changes the energies and leaves the bulk spinor overlap unchanged.

Returning to the Majorana surface with a uniform mass, the spectrum is
\begin{equation}\label{massivetilt}
 E_\pm^{(s)}=sv_Ft_xk_x\pm\sqrt{v_F^2(k_x^2+k_y^2)+m_0^2},
\end{equation}
whose minimum positive bulk excitation energy is
\begin{equation}\label{bulkgaptilt}
 E_{\rm bulk,gap}=m_0\sqrt{1-t_x^2},\qquad |t_x|<1.
\end{equation}
At $|t_x|=1$, the positive energy approaches zero as momentum tends to infinity in the linear model. For $|t_x|>1$, called type II, zero-energy bulk states occur at finite momentum. The massive and massless spectra, therefore, reach zero energy differently at the critical tilt. The negative-energy cone contribution in the gapped regime is
\begin{equation}\label{tiltchern}
 C_s=\frac{s}{2}\sgn m,\qquad
 C_s(x'<0)-C_s(x'>0)=-s.
\end{equation}
Its sign differs from Eq.~\eqref{chern} because the kinetic and mass Pauli matrices are ordered differently.

The index $s$ labels the orientation of a 2D cone. The index $i$ in Eq.~\eqref{topological} labels closed bulk Fermi surfaces. 

\subsection{Domain-wall state at arbitrary orientation}
Using the normal and tangent of Eqs.~\eqref{l-def} and \eqref{n-def}, the domain-wall Hamiltonian becomes
\begin{equation}\label{tiltHamBdG}
 H_{\rm BdG}^{(s)}=-iA^{(s)}\partial_{x'}+B^{(s)}q+m(x')\sigma_z,
\end{equation}
where $q=k_{y'}$ and
\begin{equation}\label{A-B}
\begin{aligned}
 A^{(s)}&=v_F[-s\sin\theta\,\sigma_x+\cos\theta\,\sigma_y
                      -s t_x\sin\theta\,\id],\\
 B^{(s)}&=v_F[s\cos\theta\,\sigma_x+\sin\theta\,\sigma_y
                      +s t_x\cos\theta\,\id].
\end{aligned}
\end{equation}
The normal kinetic matrix $A^{(s)}$ contains the component of the tilt perpendicular to the wall. The zero-energy equation at $q=0$,
\begin{equation}\label{TiltZeroEq}
 [-iA^{(s)}\partial_{x'}+m(x')\sigma_z]\phi_{0,s}(x')=0,
\end{equation}
admits a localized solution when
\begin{equation}\label{BoundCondition}
 Q_\theta=\sqrt{1-t_x^2\sin^2\theta}>0.
\end{equation}
The localization length is
\begin{equation}\label{XiTilt}
 \xi_t(\theta)=\frac{v_FQ_\theta}{m_0}.
\end{equation}
An explicit normalized spinor and envelope are
\begin{equation}\label{tiltspinor}
\begin{aligned}
 \phi_{0,s}(x')&=\xi_t^{-1/2}e^{-|x'|/\xi_t}\chi_s,
 \qquad\chi_s=\frac1{\sqrt2}\begin{pmatrix}1\\r_x+i r_y\end{pmatrix},\\
 r_x&=-Q_\theta\cos\theta-t_x\sin^2\theta,\\
 r_y&=s\sin\theta(t_x\cos\theta-Q_\theta).
\end{aligned}
\end{equation}
Here $r_x^2+r_y^2=1$ and $[iA^{(s)}/(v_FQ_\theta)+\sigma_z]\chi_s=0$. An overall phase can be chosen so that $\mathcal C\phi_{0,s}=\phi_{0,s}$. The components $r_x$ and $r_y$ show explicitly how tilt changes the interface spinor.

Projection of $B^{(s)}$ gives the propagation velocity
\begin{equation}\label{TiltVelocityGeneral}
 v_\parallel^{(s)}(\theta)=s v_F(t_x\cos\theta-Q_\theta).
\end{equation}
The term $s v_Ft_x\cos\theta$ is the contribution from tilt parallel to the wall. Including the transverse phase gives the wave function and dispersion
\begin{equation}\label{tiltfiniteq}
 \psi_{q,s}=e^{iqy'}e^{-iq(t_x\sin\theta/Q_\theta)x'}\phi_{0,s}(x'),
 \qquad E_s(q)=v_\parallel^{(s)}q,
\end{equation}
which solve the straight-wall linear model exactly. For $|t_x|<1$, $Q_\theta>|t_x\cos\theta|$ and $\sgn v_\parallel^{(s)}=-s$, consistently with Eq.~\eqref{tiltchern}.

The localization length is maximal when the wall normal is perpendicular to the tilt and decreases as the normal aligns with it. For $\theta=\pi/2$ or $3\pi/2$, with the mass profile in Eq.~\eqref{mass},
\begin{equation}\label{TiltVelocity}
 \xi_t=\frac{v_F}{m_0}\sqrt{1-t_x^2},\qquad
 v_\parallel^{(s)}=-s v_F\sqrt{1-t_x^2}.
\end{equation}
For a straight channel of length $L$ with antiperiodic boundary conditions, the lowest excitation energy is
\begin{equation}\label{TiltGap}
 \Delta_{\rm fs}=\frac{\pi v_F}{L}\sqrt{1-t_x^2}.
\end{equation}
For a curved closed wall, the excitation energy follows from Eq.~\eqref{loopgap}. In our angle convention, replacing $\theta$ by $\theta+\pi$ reverses both tangent and normal. Keeping $m(x')=m_0\sgn x'$ under this change also exchanges the positive- and negative-mass regions.

\begin{figure*}[t]
\centering\includegraphics[width=\linewidth]{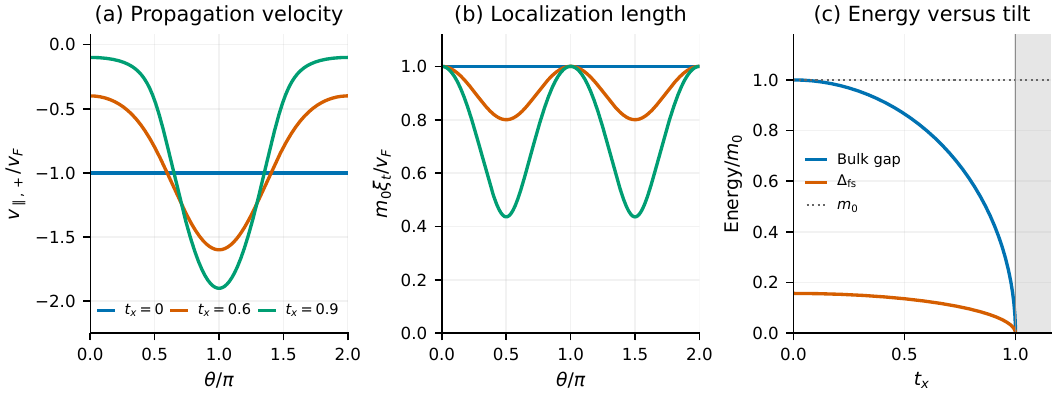}
\caption{Effect of tilt on the Majorana wall state. (a) Propagation velocity and (b) localization length versus wall angle for $s=+1$ and $t_x=0,0.6,0.9$. Negative velocity means propagation opposite to the tangent in Fig.~\ref{geometry}(b). (c) Finite-size minigap $\Delta_{\rm fs}$ and surrounding
surface bulk gap versus tilt, for a wall normal parallel
to the tilt and $L=20v_F/m_0$. Both decrease as $\sqrt{1-t_x^2}$; their ratio is $\pi/20$. The dotted line marks $m_0$. In the shaded region $t_x\geq1$, the localized solution for this wall orientation is absent. The curves follow Eqs.~\eqref{XiTilt}, \eqref{TiltVelocityGeneral}, \eqref{TiltGap}, and \eqref{bulkgaptilt}.}\label{sg-vs-tp}
\end{figure*}

For a general orientation,
\begin{equation}\label{tiltlimit}
\begin{aligned}
 \lim_{t_x\to1^-}\xi_t(\theta)&=\frac{v_F}{m_0}|\cos\theta|,\\
 \lim_{t_x\to1^-}v_\parallel^{(s)}(\theta)
 &=s v_F(\cos\theta-|\cos\theta|).
\end{aligned}
\end{equation}
The limiting values depend on orientation. For a normal parallel to the tilt, the wave function decays over a progressively shorter distance as $t_x\to1^-$. The continuum description assumes that both $1/\xi_t$ and the transverse phase wave number $|qt_x\sin\theta/Q_\theta|$ remain below the momentum cutoff. Figure~\ref{sg-vs-tp}(a,b) shows the angular dependence of
the propagation velocity and localization length at
different tilts. For a wall normal parallel to the tilt,
Fig.~\ref{sg-vs-tp}(c) shows that the finite-size minigap
and the surrounding surface bulk gap decrease by the same
factor. Their ratio therefore remains $\pi/20$ for the
parameters plotted.

For $|t_x|<1$, the surrounding surface is gapped and the wall state is localized at every angle. For $|t_x|>1$, localization requires $|t_x\sin\theta|<1$, while the 2D bulk is gapless. Translation symmetry then conserves $q$ and separates states with different momenta along the wall. The additional orientation selection found for a hexagonally warped electronic surface \cite{Vasenko2017} follows from its nonlinear momentum dependence. The present equations describe an individual cone with a linear kinetic Hamiltonian.

\section{Pairing and Majorana Interface States in Tilted Weyl Systems}\label{Weyl}

We now consider 3D Weyl bands. The superconducting spectrum depends on the nodes involved in pairing and the total momentum of each Cooper pair \cite{MengBalents,BednikZyuzinBurkov}. For example, Ref.~\cite{BednikZyuzinBurkov} finds nodal zero-momentum pairing and nodeless finite-momentum pairing in its Weyl model. Tilt also changes superconducting nodes and the momentum range of Majorana bands \cite{Alidoust}. We first calculate the singlet pairing matrix element within one Weyl band and then solve the interface problem for a spinful model with a reversing exchange field.

\subsection{Dirac bands and singlet pairing within a Weyl band}
3D Dirac semimetals have been studied theoretically and experimentally in several settings \cite{Young2012,Borisenko2014,Gibson2015,Barriga2023}. The four-component normal Hamiltonian used here is
\begin{equation}\label{tilt3d}
 H_D=v_Ft_xk_x\id_4+v_F\rho_x\otimes(\bm\sigma\cdot\bm k),
\end{equation}
where $\bm\rho$ acts on the block index and $\bm\sigma$ on the two-component spin space. Equivalently, its kinetic matrices are $\alpha_i=\rho_x\otimes\sigma_i$. The eigenvalues are
\begin{equation}\label{diracdoublet}
 E_{\rho,h}=v_Ft_xk_x+\rho h v_F|\bm k|,\qquad \rho,h=\pm1,
\end{equation}
with $h$ the spin helicity. Each energy branch contains two degenerate states. The conserved label $\rho_x=\pm1$ separates the Hamiltonian into two Weyl blocks with opposite Chern numbers \cite{Armitage}. Their total Chern number is zero. We use this block label to select the state whose pairing matrix element is evaluated below.

At charge neutrality, the massless zero-energy set obeys
\begin{equation}\label{weylzeroset}
 (t_x^2-1)k_x^2=k_y^2+k_z^2.
\end{equation}
As in Eq.~\eqref{DiracPoint3d}, type I has a zero-energy point. For $|t_x|>1$, the linear dispersion produces open electron and hole sheets, whose shapes at larger momenta depend on band curvature \cite{Soluyanov}.

Within the $\rho_x=+1$ block, the lower-band spinor is
\begin{equation}\label{weylspinor}
 |u_-(\bm k)\rangle=\frac1{\sqrt2}
 \begin{pmatrix}
 -\sin(\vartheta/2)e^{-i\varphi}\\\cos(\vartheta/2)\\
 -\sin(\vartheta/2)e^{-i\varphi}\\\cos(\vartheta/2)
 \end{pmatrix},
\end{equation}
where $\vartheta,\varphi$ are spherical momentum angles, distinct from the wall angle $\theta$. With $\bm{\mathcal A}=i\langle u_-|\nabla_{\bm k}u_-\rangle$, one obtains
\begin{equation}\label{weylberry}
\begin{gathered}
 \mathcal A_k=\mathcal A_\vartheta=0,\qquad
 \mathcal A_\varphi=\frac{1-\cos\vartheta}{2k\sin\vartheta},\\
 \mathcal F_{\vartheta\varphi}=\frac12\sin\vartheta,
 \qquad \bm\Omega=\frac{\hat{\bm k}}{2k^2}.
\end{gathered}
\end{equation}
Here $\mathcal A_\varphi$ is the component along the azimuthal unit vector, and $\mathcal F_{\vartheta\varphi}$ is the curvature integrated with $d\vartheta\,d\varphi$. The Berry flux of the selected lower band gives
\begin{equation}\label{WeylNode}
 C_W=\frac1{2\pi}\oint_{S^2}\bm\Omega\cdot d\bm S=1.
\end{equation}
The identity tilt leaves the spinor and its Berry flux unchanged. The quantized flux in Eq.~\eqref{WeylNode} is defined
within the conserved $\rho_x=+1$ block, whose lower band
is nondegenerate on a closed surface enclosing but not
passing through the node. In type II, both energy branches can cross the Fermi energy, so their occupation must also be included when calculating a response \cite{WiederBJ,Udagawa,tiwari}. The sign convention in Eq.~\eqref{WeylNode} uses $\bm{\mathcal A}=i\langle u|\nabla_{\bm k}u\rangle$, as specified above.

For a real intra-block spin-singlet amplitude $\Delta_s$, use
\begin{equation}\label{weylpairmatrix}
 \hat\Delta=\Delta_s\rho_0\otimes i\sigma_y.
\end{equation}
Momentum inversion sends $(\vartheta,\varphi)$ to $(\pi-\vartheta,\varphi+\pi)$, so
\begin{equation}\label{weylpairaction}
 \hat\Delta|u_-^*(-\bm k)\rangle=\frac{\Delta_s}{\sqrt2}
 \begin{pmatrix}
 \sin(\vartheta/2)\\-\cos(\vartheta/2)e^{i\varphi}\\
 \sin(\vartheta/2)\\-\cos(\vartheta/2)e^{i\varphi}
 \end{pmatrix},
\end{equation}
and the projected amplitude is
\begin{equation}\label{p-amplitude}
 \Delta_{\rm eff}(\bm k)=\langle u_-(\bm k)|\hat\Delta|u_-^*(-\bm k)\rangle
 =-\Delta_s e^{i\varphi}.
\end{equation}
The phase $e^{i\varphi}=(k_x+i k_y)/\sqrt{k_x^2+k_y^2}$ follows from the Bloch-spinor convention in Eq.~\eqref{weylspinor}. As in the 2D singlet projection \cite{FuKane,ReadGreen}, the phase winds with momentum. In three dimensions, separate smooth spinor conventions are needed near the two poles of the momentum sphere. The projected amplitude has magnitude $\Delta_s$, whereas an ordinary $p_x+i p_y$ function has magnitude proportional to $\sqrt{k_x^2+k_y^2}$. The topology of Cooper pairing on Fermi surfaces with nonzero Chern numbers is discussed in Ref.~\cite{Haldane2018}.

The single-band approximation assumes that the energy separation from the other band is much larger than the pair potential. For a node at crystal momentum $\bm K$, intra-node pairing of $\bm K\pm\bm k$ has center-of-mass momentum $2\bm K$ modulo a reciprocal vector; uniform pairing can instead join different nodes. The pairing momentum is therefore an essential part of the superconducting Hamiltonian \cite{BednikZyuzinBurkov}. We specify it explicitly for the wall calculation below.

\subsection{Exchange-wall BdG Hamiltonian and zero-momentum states}
For a plane wall at $z=0$, take the electron Hamiltonian and singlet pair potential
\begin{equation}\label{electron3d}
\begin{aligned}
 h_e(\bm k,z)&=v_F\bm\sigma\cdot\bm k+v_Ft_xk_x\id
                  +b(z)\sigma_x-\mu\id,\\
 \hat\Delta_e&=\Delta_0 i\sigma_y,\qquad b(z)=b_0\sgn z.
\end{aligned}
\end{equation}
Here $b_0>0$ is an exchange or Zeeman scale. Momentum is measured from a continuum origin identified with a time-reversal-invariant crystal momentum, so the stated uniform singlet term pairs $\bm k$ and $-\bm k$. At a node with crystal momentum $\bm K$, the corresponding intra-node pairing instead carries momentum $2\bm K$; uniform inter-node pairing requires explicit valley degrees of freedom. In the conventional electron--hole basis the diagonal blocks are $h_e(\bm k,z)$ and $-h_e^{\mathrm T}(-\bm k,z)$. Transforming the hole components by $i\sigma_y$ gives, in the basis
$\Psi_{\bm k}=(c_{\bm k\uparrow},c_{\bm k\downarrow},c^\dagger_{-\bm k\downarrow},-c^\dagger_{-\bm k\uparrow})^{\mathrm T}$,
\begin{equation}\label{H_BdG_3D}
\begin{aligned}
 \mathcal H={}&v_F\bm\sigma\cdot\bm k\,\tau_z+v_Ft_xk_x\id_4\\
 &+b(z)\sigma_x+\Delta_0\tau_x-\mu\tau_z.
\end{aligned}
\end{equation}
The Pauli matrices $\bm\tau$ now act on Nambu space. We choose the global gauge $\Delta_0>0$. The intrinsic particle-hole operator is $\mathcal C=\sigma_y\tau_y\mathcal K$, with $\mathcal C^2=+1$. The momentum-even exchange term has no $\tau_z$, while the odd identity tilt has no $\tau_z$ either, as follows directly from the electron--hole construction.

The field $b\sigma_x$ shifts the normal-state Weyl node along $k_x$. Its sign reversal defines an \emph{exchange wall with uniform pairing}. The symbols $b$ and $m$ distinguish this exchange field from the surface mass of Sections~\ref{Anis} and \ref{Tilt}. Equation~\eqref{H_BdG_3D} keeps both electron-spin components and follows directly from Eq.~\eqref{electron3d}. The relation between Weyl-node structure and superconducting boundary states is reviewed in Ref.~\cite{Armitage}; proximity-induced Weyl-surface states are treated in Refs.~\cite{chen2016weyl,chen2017josephson}.

At $k_x=k_y=\mu=0$, the zero-mode equation is
\begin{equation}\label{zero_mode_3D}
 [-iv_F\sigma_z\tau_z\partial_z+b(z)\sigma_x+\Delta_0\tau_x]\Phi=0,
\end{equation}
or
\begin{equation}\label{JR_3D}
\begin{gathered}
 \relax[v_F\partial_z-b(z)M_1-\Delta_0M_2]\Phi=0,\\
 M_1=\sigma_y\tau_z,\qquad M_2=\sigma_z\tau_y.
\end{gathered}
\end{equation}
Since $[M_1,M_2]=0$, use normalized simultaneous eigenstates with $M_1\chi_\eta=-\chi_\eta$ and $M_2\chi_\eta=\eta\chi_\eta$:
\begin{equation}\label{spinor3d0}
 \chi_\eta(0)=\frac12\begin{pmatrix}1\\-i\\i\eta\\-\eta\end{pmatrix},\qquad \eta=\pm1.
\end{equation}
For $b_0>\Delta_0$, the normalized states are
\begin{equation}\label{wf_3D}
 \Phi_\eta(z)=\sqrt{\frac{b_0^2-\Delta_0^2}{v_Fb_0}}
 e^{-b_0|z|/v_F+\eta\Delta_0z/v_F}\chi_\eta(0).
\end{equation}
The envelopes satisfy $\int dz\,\Phi_\eta^\dagger\Phi_\eta=1$. Since $\mathcal C\chi_\eta=\eta\chi_\eta$, the states $\Phi_+$ and $i\Phi_-$ are self-conjugate at zero momentum. The localization lengths on the right and left are
\begin{equation}\label{xi3d}
 \xi_{\eta,R}=\frac{v_F}{b_0-\eta\Delta_0},\qquad
 \xi_{\eta,L}=\frac{v_F}{b_0+\eta\Delta_0}.
\end{equation}
Pairing changes the exponential decay on each side of the wall. For $\Delta_0\ll b_0$, both localization lengths approach $v_F/b_0$. Spatial control of Majorana states through inhomogeneous superconducting parameters has also been studied in Ref.~\cite{marra2017controlling}.

\subsection{Exact finite-momentum states and surface dispersion}
At $\mu=0$, the solution extends to finite $k_x$ by defining
\begin{equation}\label{Gdef}
 G=v_Fk_x\sigma_y+\Delta_0M_2,\qquad R=\sqrt{v_F^2k_x^2+\Delta_0^2}.
\end{equation}
The identities $[G,M_1]=0$ and $G^2=R^2\id_4$ give normalized spinors
\begin{equation}\label{spinor3dk}
 \chi_\eta(k_x)=\frac{(R+\eta G)\chi_\eta(0)}{\sqrt{2R(R+\Delta_0)}}.
\end{equation}
They satisfy $G\chi_\eta=\eta R\chi_\eta$ and $M_1\chi_\eta=-\chi_\eta$. Substitution into the full differential equation yields
\begin{align}
 \Phi_{\eta,k_x}(z)&=\sqrt{\frac{b_0^2-R^2}{v_Fb_0}}
 e^{-b_0|z|/v_F+\eta Rz/v_F}\chi_\eta(k_x),\label{finite3d}\\
 E_\eta(k_x,k_y)&=v_F(t_xk_x-k_y),\label{surface_disp}\\
 |k_x|&<k_c=\frac{\sqrt{b_0^2-\Delta_0^2}}{v_F}.\label{arcrange}
\end{align}
Both localized states have the same dispersion. The $k_y$ term acts as an identity shift because $\sigma_y\tau_z=-\id$ in this subspace. The transverse spinor and envelope depend on $k_x$ through $G$ and $R$. Their localization lengths are $v_F/(b_0\mp\eta R)$ on the right and left. One length diverges as $|k_x|\to k_c$, where the state spreads into the bulk. Appendix~\ref{app:numerics} checks the energies and the two localized wave functions by direct diagonalization.

\begin{figure*}[t]
\centering\includegraphics[width=\linewidth]{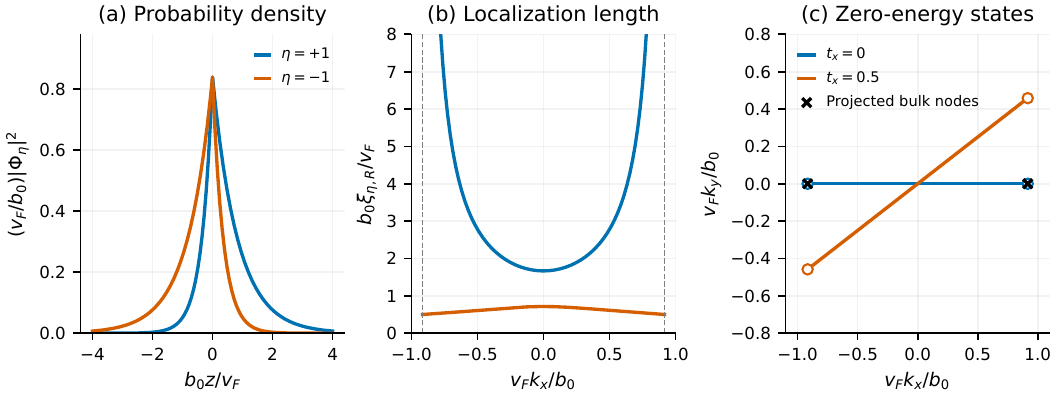}
\caption{Interface states at an exchange wall for $\mu=0$ and $\Delta_0/b_0=0.4$. (a) Normalized probability densities perpendicular to the wall at $k_x=0$. The two states, $\eta=\pm1$, have opposite left-right asymmetry. (b) Localization lengths in the $z>0$ region versus momentum along the wall. Interchanging $\eta$ gives the lengths for $z<0$. The dashed lines mark $k_x=\pm k_c$, where one localization length diverges. (c) Momenta of zero-energy interface states for $t_x=0$ and $0.5$. Open circles mark the endpoints at which the states spread into the bulk. Crosses mark the projected bulk nodes, where two bands meet. For nonzero tilt, these nodes have nonzero energies, while the interface endpoints join zero-energy bulk states at $k_y=t_xk_x$. Length and energy units are $v_F/b_0$ and $b_0$.}\label{figweyl}
\end{figure*}

For either uniform exchange domain, the $\mu=0$ bulk spectrum is
\begin{equation}\label{bulk3d}
\begin{aligned}
 E_{\alpha,\beta}(\bm k)&=v_Ft_xk_x\\
 &\quad+\alpha\sqrt{v_F^2(k_y^2+k_z^2)+(b_0+\beta R)^2},
\end{aligned}
\end{equation}
where $\alpha,\beta=\pm1$. It has band-touching points at $(\pm k_c,0,0)$ with energies $\pm v_Ft_xk_c$. The bulk therefore remains gapless in the parameter range supporting these interface states, even for type-I tilt. The zero-energy interface segment is
\begin{equation}\label{zeroarc}
 k_y=t_xk_x,\qquad |k_x|<k_c.
\end{equation}
For $t_x=0$ its endpoints coincide with the projected nodes. For nonzero tilt, its endpoints $(\pm k_c,\pm t_xk_c)$ instead merge with zero-energy bulk states at those momenta; the projected band-touching points remain at $k_y=0$ and have nonzero energies. Figure~\ref{figweyl}(a) shows the opposite left--right
asymmetry of the two zero-momentum states, while
Fig.~\ref{figweyl}(b) shows the divergence of one
localization length as $|k_x|\to k_c$.
Figure~\ref{figweyl}(c) distinguishes the endpoints of
the zero-energy interface segment from the projected
bulk nodes at nonzero tilt.

For small chemical potential, projection about $k_x=k_y=\mu=0$ must include the spatial overlap as well as the spinor matrix elements. With $P$ the projector onto $M_1=-1$, one has $P\sigma_x\tau_zP=0$ and $P\sigma_y\tau_zP=-P$. In the phase convention of Eq.~\eqref{spinor3d0}, $\tau_z$ exchanges the two spinors, while the overlap of their different normalized envelopes is
\begin{equation}\label{overlap}
 S=\frac{b_0^2-\Delta_0^2}{v_Fb_0}\int dz\,e^{-2b_0|z|/v_F}
 =1-\frac{\Delta_0^2}{b_0^2}.
\end{equation}
Consequently,
\begin{equation}\label{H_surf_final}
\begin{aligned}
 H_{\rm surf}^{(1)}&=v_F(t_xk_x-k_y)\id-\mu S\widetilde\tau_x,\\
 E_\pm^{(1)}&=v_F(t_xk_x-k_y)\pm|\mu|S.
\end{aligned}
\end{equation}
Here $\widetilde{\tau}_x$ acts in the interface-state basis
$(\Phi_+,\Phi_-)$ of Eq.~\eqref{wf_3D}.
Equation~\eqref{H_surf_final} is valid to first order in
$v_Fk_x$, $v_Fk_y$, and $\mu$, with their magnitudes small
compared with $b_0-\Delta_0$, the minimum bulk excitation
energy at $k_x=k_y=0$ in the unperturbed problem.
This separation is defined at fixed parallel momentum;
the three-dimensional bulk remains gapless at the nodes.
At $\mu=0$, the full momentum dependence and localization
interval are given by Eqs.~\eqref{finite3d}--\eqref{zeroarc}.

\subsection{Superconducting gap in a projected band}
To examine how tilt affects the superconducting gap after band projection, consider an upper band with normal energy $v_F|\bm k|-\mu$ and a pair amplitude of constant magnitude. In a region of momentum space where the Bloch spinor is smooth, its two-component Nambu Hamiltonian is
\begin{equation}\label{H_effective_bulk}
 H_{\rm loc}=\bm h(\bm k)\cdot\bm\tau+v_Ft_xk_x\id,
\end{equation}
where
\begin{equation}\label{h_vector}
 \bm h=\left(\Delta_0\frac{k_x}{k_\perp},
              -\Delta_0\frac{k_y}{k_\perp},v_F|\bm k|-\mu\right),
 \qquad k_\perp=\sqrt{k_x^2+k_y^2}.
\end{equation}
Equation~\eqref{H_effective_bulk} describes the uniform projected band, with no exchange wall. The displayed spinor convention is defined for $k_\perp>0$; near the polar axis a different Bloch-spinor phase convention is required. A Berry-curvature calculation uses the full BdG eigenstate, including the momentum dependence of this Bloch basis. The relation $H_{\rm loc}(-\Delta_0)=\tau_zH_{\rm loc}(\Delta_0)\tau_z$ also shows that changing the sign of a uniform pair amplitude is a unitary change of basis.

The zero-energy condition is
\begin{equation}\label{pockets}
 (v_Ft_xk_x)^2=\Delta_0^2+(v_F|\bm k|-\mu)^2.
\end{equation}
For $\mu>0$, minimization over momentum gives the gap-closing threshold
\begin{equation}\label{projectedgapthreshold}
 |t_x|\geq t_c=\frac{\Delta_0}{\sqrt{\mu^2+\Delta_0^2}}<1,
\end{equation}
attained along the tilt axis at $v_F|k_x|=(\mu^2+\Delta_0^2)/\mu$. This minimum describes the projected band when that momentum lies within its range of validity and the separation from other bands exceeds the pair potential. The superconducting gap in this model closes at $t_c<1$, while the normal-state cone is still type I.

The superconducting gap also determines which response theory applies. The bulk phase action in Eq.~\eqref{phaseaction} describes the fully gapped parent of Appendix~\ref{app:parent}, with integer Chern numbers defined on closed normal-state Fermi surfaces. The electromagnetic response of a normal Weyl semimetal is described by a different action \cite{Zyuzin2012}. In the surface problem studied here, the connection to the superconducting phase theory is the mass produced by the relative pairing phase.

\section{Conclusions}\label{summ}
We have obtained analytical wave functions, propagation velocities, and localization lengths for Majorana states at anisotropic and tilted surface mass walls. For an invertible velocity tensor, its determinant fixes the propagation direction at a specified mass wall, while the wall angle controls the speed and localization length. The product of the propagation speed and localization length is independent of angle. As one principal velocity tends to zero, the interface speed approaches zero at most fixed orientations but remains finite when the wall normal lies along the direction of the vanishing principal velocity.

Cone tilt affects propagation and localization through different components. The component normal to the wall changes the localization length, and the parallel component contributes directly to the propagation velocity. When the normal is parallel to the tilt, the localization length, propagation speed, and lowest finite-size excitation energy decrease together toward the critical tilt. For a smooth closed wall, the lowest positive level with
antiperiodic boundary conditions is fixed by the propagation
time around the perimeter. A fully gapped 3D superconducting parent
provides the two-dimensional surface Hamiltonian. The
relative phase between its triplet and singlet pairing
components generates the mass, whose sign reversal supports
the one-dimensional channel.

In the separate 3D Weyl model, a planar
exchange wall supports 2D interface states
with equal dispersion at zero chemical potential, while
the bulk remains gapless. Pairing determines their different decay lengths on the two sides and the finite interval of allowed momentum along the wall. Tilt shifts their energies and changes the line of zero-energy interface states. A small chemical potential splits the two branches by an amount set by the overlap of their transverse wave functions. Direct diagonalization confirms the analytical energies and wave functions as the grid is refined and the simulation box is enlarged. Together, these results give explicit relations between the velocity tensor, tilt, pairing amplitude, wall orientation, and the propagation and confinement of Majorana interface states.

\section*{Acknowledgments}
We acknowledge support from the Research Council of Norway through Grant Nos. 353919 and 361800 ``QTransMag'', and Grant No. 262633 ``QuSpin''.

\emph{Disclosure of generative AI use.} During the preparation and revision of this manuscript, the authors used ChatGPT to assist with scientific discussions, manuscript editing, and debugging of numerical code. The tools were directed using the authors’ manuscript, model definitions, and specific revision requests. The authors take full responsibility for the accuracy, integrity, and conclusions of this work.

\section*{Data Availability}
Data supporting the findings of this study are contained
within the article and its appendices.

\appendix
\section{A gapped parent and its Majorana surface projection}\label{app:parent}
We derive the two-component surface Hamiltonian of Section~\ref{Anis} from a spinful electron Hamiltonian $h_p(\bm k)=M(\bm k)\id$, with $M(\bm k)=B|\bm k|^2-\mu_p$, $B>0$, and a triplet pair vector
\begin{equation}\label{parentd}
 \bm d_p(\bm k)=(-d_z,d_x,v_zk_z),\qquad v_z>0,
\end{equation}
where $d_x=v_{xx}k_x+v_{yx}k_y$ and $d_z=v_{xz}k_x+v_{yz}k_y$. Add a small imaginary spin-singlet component $-i\delta_s$, so that the electron pair matrix is
\begin{equation}\label{parentpair}
 \hat\Delta_p=(\bm d_p\cdot\bm\sigma-i\delta_s\id)i\sigma_y.
\end{equation}
The odd triplet vector and even singlet term obey $\hat\Delta_p(\bm k)=-\hat\Delta_p^{\mathrm T}(-\bm k)$. In the time-reversed-hole Nambu basis used in Section~\ref{Weyl}, the BdG operator is
\begin{equation}\label{parentH}
 \mathcal H_p=M\tau_z+\tau_x\bm d_p\cdot\bm\sigma+\delta_s\tau_y.
\end{equation}
It has $\mathcal C=\sigma_y\tau_y\mathcal K$. At $\delta_s=0$, it also has $\mathcal T=i\sigma_y\mathcal K$ and is an anisotropic representative of the familiar fully gapped triplet parent \cite{QiZhang,xiaozhang}. For $\mu_p>0$ and invertible $\bm V$, its spectrum $E^2=M^2+|\bm d_p|^2$ is gapped, and the quadratic term specifies its behavior at large momentum. The negative-$\mu_p$ vacuum is topologically trivial. 

To obtain the boundary states in the linear approximation, take $M(z)=M_0\sgn z$ and set $k_x=k_y=\delta_s=0$. The zero-mode equation is
\begin{equation}\label{parentwall}
 [v_z\partial_z+M(z)\tau_y\sigma_z]\Phi=0.
\end{equation}
The decaying states have $\tau_y\sigma_z=+1$ and envelope $f(z)=\sqrt{M_0/v_z}\,e^{-M_0|z|/v_z}$. The quadratic term changes the short-distance shape of this linear-model envelope while preserving the two surface components. In the Nambu order stated in Section~\ref{Weyl}, introduce
\begin{equation}\label{parentspinors}
\begin{gathered}
 u_1=\frac1{\sqrt2}\begin{pmatrix}1\\0\\i\\0\end{pmatrix},\quad
 u_2=\frac1{\sqrt2}\begin{pmatrix}0\\1\\0\\-i\end{pmatrix},\\
 W=\frac1{\sqrt2}\bigl(u_1-i u_2,\,i u_1-u_2\bigr).
\end{gathered}
\end{equation}
Both columns of $W$ are self-conjugate under $\mathcal C$. Writing $\widetilde\sigma_i$ for Pauli matrices in this surface basis, direct projection gives
\begin{equation}\label{parentprojection}
\begin{aligned}
 W^\dagger\tau_x\sigma_y W&=\widetilde\sigma_x,&
 W^\dagger\tau_x\sigma_x W&=-\widetilde\sigma_z,\\
 W^\dagger\tau_y W&=-\widetilde\sigma_y,&
 W^\dagger W&=\id.
\end{aligned}
\end{equation}
Consequently, to leading order in surface momenta and the singlet perturbation,
\begin{equation}\label{parentsurface}
\begin{aligned}
 H_{\rm surf}&=d_x\widetilde\sigma_x+d_z\widetilde\sigma_z
                  +m(x')\widetilde\sigma_y,\\
 m(x')&=-\int dz\,|f(z)|^2\delta_s(x',z).
\end{aligned}
\end{equation}
Dropping the tildes reproduces the surface Hamiltonian of Section~\ref{Anis}, with $\mathcal C=\mathcal K$ on its real Majorana field. Opposite imaginary singlet components correspond to opposite relative superconducting phases and give the mass-sign wall used there. This realizes the surface mechanism discussed in Sec.~IV.B of Ref.~\cite{xiaozhang} while making the Nambu origin explicit.

Adding the term $(\bm w\cdot\bm k_\parallel)\id_4$ produces a tilt of the surface spectrum. It preserves particle-hole symmetry and breaks time reversal. For the parent Hamiltonian above at $\delta_s=0$, the spectrum is
\begin{equation}\label{parenttiltspectrum}
 E_\pm=\bm w\cdot\bm k_\parallel\pm
 \sqrt{M^2+|\bm V\bm k_\parallel|^2+v_z^2k_z^2}.
\end{equation}
A sufficient condition for its gap to remain open is
\begin{equation}\label{parenttiltbound}
 \bm w^{\mathrm T}(\bm V^{\mathrm T}\bm V)^{-1}\bm w<1.
\end{equation}
Indeed, the weighted Cauchy--Schwarz inequality bounds $(\bm w\cdot\bm k_\parallel)^2$ by the left-hand side of Eq.~\eqref{parenttiltbound} times $|\bm V\bm k_\parallel|^2$. With $\mu_p>0$, the two terms in Eq.~\eqref{parenttiltspectrum} cannot cancel, and the quadratic regularization controls large momentum. For isotropic in-plane velocities and $\bm w=s v_Ft_x\hat{\bm x}$, this condition is $|t_x|<1$. Thus this parent remains gapped for the type-I tilted surface. The tilt breaks time-reversal symmetry, and the singlet phase is chosen to produce the mass profile used in Sections~\ref{Anis} and \ref{Tilt}.

\section{Numerical verification of the interface states}\label{app:numerics}
We test the anisotropic, tilted, and Weyl wall solutions by direct numerical diagonalization of their differential operators. The transverse coordinate is placed on a periodic interval of length $L_\perp$. For an odd number $N$ of equally spaced points, the Fourier differentiation matrix is
\begin{equation}\label{spectralD}
 D_{jl}=\begin{cases}
 \dfrac{\pi}{L_\perp}(-1)^{j-l}
 \csc\!\left[\dfrac{\pi(j-l)}{N}\right],&j\ne l,\\
 0,&j=l.
 \end{cases}
\end{equation}
The Hermitian matrix is assembled as
\begin{equation}\label{spectralH}
 H_N=-iD\otimes A+\id_N\otimes B_0
       +\operatorname{diag}(\sgn x_j)\otimes M_0,
\end{equation}
where $A$ is the transverse kinetic matrix, $B_0$ contains the constant tangential-momentum and pairing terms, and $M_0$ is the wall matrix, including its amplitude. The grid contains $x_j=0$, where the sign profile is assigned its mean value zero. A periodic sign profile necessarily includes a second, oppositely oriented wall at the box boundary. We select the states with the largest probability in $|x|<8$, using the units given below. Increasing $L_\perp$ then tests the influence of the second wall.

\begin{figure*}[t]
\centering\includegraphics[width=.95\linewidth]{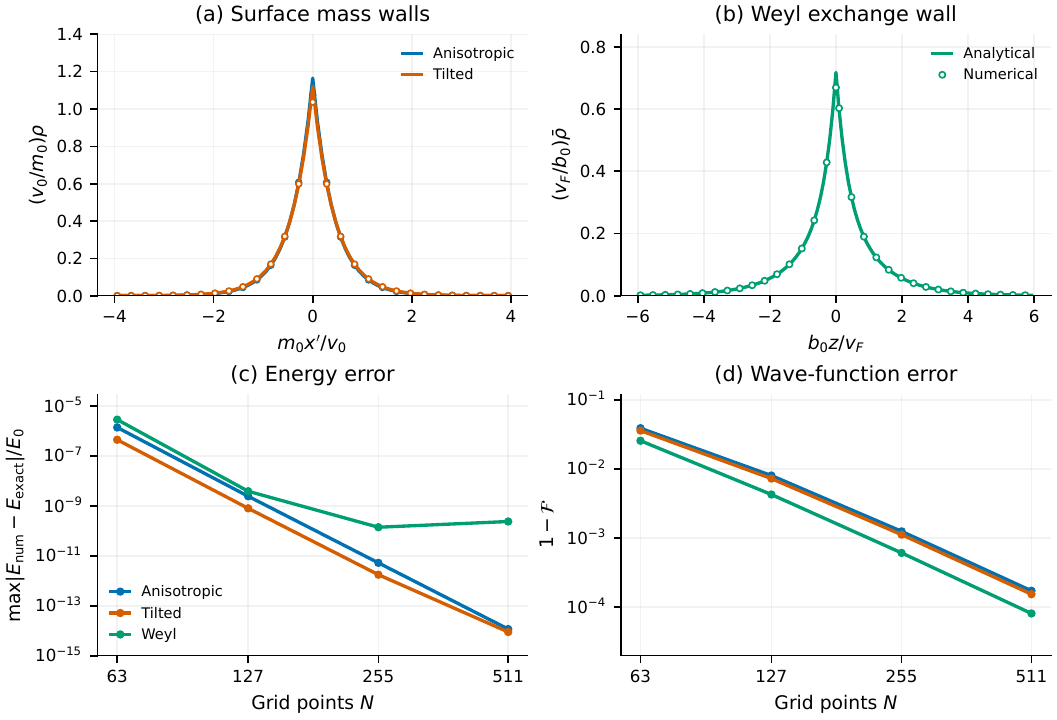}
\caption{Numerical verification of the analytical interface states. Parameters are given in Appendix~\ref{app:numerics}. (a) Probability densities perpendicular to the anisotropic and tilted surface mass walls. Solid lines are analytical solutions on the infinite line; open circles are numerical results for $N=511$ grid points in a box of length $L_\perp=48$. The tilt calculation uses $v_0=v_F=1$. (b) Mean probability density of the two Weyl exchange-wall states, $\bar\rho=\frac12\sum_\eta\Phi_\eta^\dagger\Phi_\eta$. Averaging over both states gives the same density for any orthonormal combinations of the degenerate eigenvectors. Only selected numerical grid points are marked in (a,b). (c) Maximum energy error and (d) wave-function error $1-\mathcal F$, defined by Eq.~\eqref{subspacefidelity}, versus the number of grid points. Zero error corresponds to exact agreement. The energy unit is $E_0=m_0$ for the surface mass walls and $E_0=b_0$ for the Weyl wall. The Weyl energy error at the finest grids is set by the second wall at the periodic box boundary; increasing the box length reduces this error, as shown in Appendix~\ref{app:numerics}.}\label{fignumerics}
\end{figure*}

The tests use the following dimensionless parameters. For anisotropy, $v_0=m_0=1$, $\bm V=W$ of Fig.~\ref{anisotropic}, $\theta=0.63$, and $q=0.25$. For tilt, $v_F=m_0=1$, $s=+1$, $t_x=0.6$, $\theta=0.85$, and $q=0.25$. For the Weyl wall, $v_F=b_0=1$, $\Delta_0=0.4$, $t_x=0.4$, $k_x=0.35$, $k_y=0.18$, and $\mu=0$. In each model $L_\perp=48$ and $N=63,127,255,511$. The exact energies are $0.3146232354$, $-0.1241627832$, and $-0.04$, respectively; the Weyl energy is twofold degenerate.

We compare both the energies and the wave functions. Let the columns of $U_{\rm a}$ and $U_{\rm n}$ be the normalized analytical and numerical bound states on the grid. Their overlap is measured by
\begin{equation}\label{subspacefidelity}
 \mathcal F=\frac1r\|U_{\rm a}^\dagger U_{\rm n}\|_{\rm F}^2,
\end{equation}
where $r=1$ for either two-component wall and $r=2$ for the Weyl wall. The squared norm is the sum of the squared absolute values of all matrix elements. We orthonormalize the analytical states on the grid before taking this overlap. For the two degenerate Weyl states, Eq.~\eqref{subspacefidelity} compares the spaces they span, independently of the particular linear combinations returned by the eigensolver. Identical spaces give $\mathcal F=1$, so we use $1-\mathcal F$ as the wave-function error. Figure~\ref{fignumerics}(a,b) shows agreement between
the analytical and numerical probability densities.
Panels (c,d) show that the wave-function error decreases
with grid refinement, while the Weyl energy error
approaches a floor set by the finite box.
Table~\ref{tab:numerics} lists the errors at the finest grid.

\begin{table}[tb]
\caption{Errors at $N=511$ and $L_\perp=48$. The energy unit $E_0$ is $m_0$ for the two-component models and $b_0$ for the Weyl model. Energies are much more accurate than the spatial profiles; both are checked.}\label{tab:numerics}
\begin{ruledtabular}
\begin{tabular}{lcc}
Model & $\max|E_{\rm n}-E_{\rm a}|/E_0$ & $1-\mathcal F$\\
Anisotropic & $1.2\times10^{-14}$ & $1.7\times10^{-4}$\\
Tilted & $8.9\times10^{-15}$ & $1.6\times10^{-4}$\\
Weyl & $2.4\times10^{-10}$ & $8.2\times10^{-5}$
\end{tabular}
\end{ruledtabular}
\end{table}

For the Weyl state, the energy error saturates with grid refinement because of the remote periodic wall. At nearly fixed spacing $\delta z\simeq0.094$, increasing $L_\perp$ from $32$ to $48$ to $64$ reduces this error from $4.4\times10^{-7}$ to $2.4\times10^{-10}$ to $2.7\times10^{-13}$, while the wave-function error $1-\mathcal F$ stays near $8.2\times10^{-5}$. The remaining wave-function error is therefore controlled mainly by the grid spacing in this comparison. As a further check, at $k_x=0$, $k_y=0.12$, $\Delta_0=0.4$, $N=511$, and $L_\perp=48$, the numerical splitting divided by $2\mu$ is $0.83901$, $0.83947$, and $0.83958$ for $\mu=0.04,0.02,0.01$. These values agree within $10^{-3}$ with the first-order spatial-overlap factor $S=0.84$ in Eq.~\eqref{overlap}; finite-grid and higher-order corrections are retained in this comparison. 

\clearpage
\bibliography{bib}

\end{document}